\documentclass[aip,reprint, nofootinbib]{revtex4-1}  % for review and submission

\usepackage{silence}
\usepackage{graphicx}  % needed for figures
\usepackage{dcolumn}   % needed for some tables
\usepackage{bm}        % for math
\usepackage{amssymb}   % for math
\usepackage{float}
\usepackage{epsfig,amsmath,amssymb,color,float,setspace}
\usepackage[utf8]{inputenc}
\usepackage[T1]{fontenc}
\usepackage{titlesec}
\usepackage{booktabs}
\usepackage{gensymb}
\usepackage[symbol]{footmisc}

\begin{document}
\title{Structural and electronic properties of the spin-filter material CrVTiAl with disorder}
\author{Gregory M. Stephen*}
\affiliation{Department of Physics, Northeastern University, Boston, MA 02115 USA}\footnote{g.stephen@northeastern.edu}
\author{Gianina Buda}
\affiliation{Department of Physics, Northeastern University, Boston, MA 02115 USA}
\author{Michelle E. Jamer}
\affiliation{NIST Center for Neutron Research, National Institute of Standards and Technology, Gaithersburg, MD 20899, USA}
\affiliation{Department of Physics, U.S. Naval Academy, Annapolis, MD 21402 USA}
\author{Christopher Lane}
\affiliation{Department of Physics, Northeastern University, Boston, MA 02115 USA}
\author{Stanislaw Kaprzyk}
\affiliation{Faculty of Physics and Applied Computer Science, AGH University of Science and Technology, aleja Mickiewicza 30, 30-059 Krakow, Poland}
\affiliation{Department of Physics, Northeastern University, Boston, MA 02115 USA}
\author{Bernardo Barbiellini}
\affiliation{ Department of Physics, School of Engineering Science, Lappeenranta University of Technology, FI-53851 Lappeenranta, Finland}
\affiliation{Department of Physics, Northeastern University, Boston, MA 02115 USA}
\author{David Graf}
\affiliation{National High Magnetic Field Laboratory, Florida State University, Tallahassee, Florida 32310, USA}
\author{Laura H. Lewis}
\affiliation{Department of Chemical Engineering, Northeastern University, Boston, MA 02115 USA}
\affiliation{Department of Mechanical and Industrial Engineering, Northeastern University, Boston, MA 02115 USA}
\author{Arun Bansil}
\affiliation{Department of Physics, Northeastern University, Boston, MA 02115 USA}
\author{Don Heiman}
\affiliation{Department of Physics, Northeastern University, Boston, MA 02115 USA}
\date{\today}
\begin{abstract}

The effects of chemical disorder on the transport properties of the spin-filter material CrVTiAl are investigated experimentally and theoretically. Synchrotron X-ray diffraction experiments on bulk CrVTiAl and the associated Rietveld analysis indicate that the crystal structure consists primarily of a mixture of a partially ordered B2 phase, a fully disordered A2 phase and a small component of an ordered L2\textsubscript{1} or Y phase. High temperature resistivity measurements confirm the existence of a band gap. First-principles, all-electron, self-consistent electronic structure computations show that the chemically disordered A2 and B2 phases are metallic, while the spin-filter properties of the ideal Y-type phase are preserved in the presence of L2\textsubscript{1} disorder. The Hall coefficient is found to decrease with increasing temperature, similar to the measured increase in the conductivity, indicating the presence of thermally activated semiconductor-like carriers.

\end{abstract}

 \maketitle
\section{Introduction}
\begin{figure}[t!]
\includegraphics[width=0.32\textwidth]{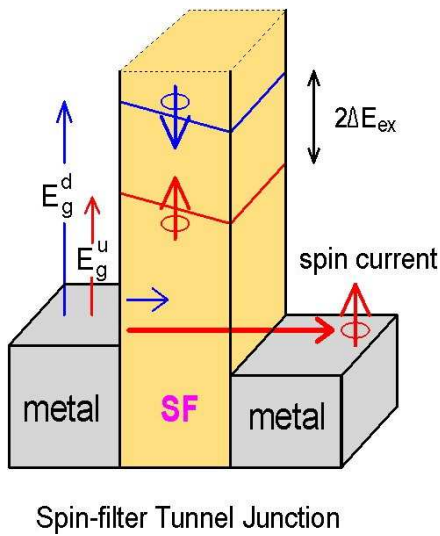}
\caption{Illustration of an essentially \textit{nonmagnetic} spin-filter device composed of a spin-filter tunneling barrier sandwiched between two nonmagnetic electrodes. Biasing the device allows spin-up electrons to tunnel through, while the higher barrier for spin-down electrons prohibits those electrons from tunneling.}
\label{fig:SF_device}
\end{figure}

Spin-filter materials (SFMs) are promising candidates for generating spin-polarized currents that could be useful for producing next-generation memory and logic devices. \cite{Graf2016a,Wolf2001} One of the advantages of a spin-filter device is that it requires only nonmagnetic contacts, as opposed to a standard magnetic tunnel junction (MTJ) that uses ferromagnetic (FM) contacts. SFMs are semiconductors in which a magnetic exchange interaction splits the spin degeneracy, leading to a larger bandgap for one spin direction.\cite{Moodera2010} The exchange splitting energy 2$\Delta$E\textsubscript{ex} allows an SFM to be used as a tunneling barrier to filter out one spin direction as illustrated in Fig. \ref{fig:SF_device}. As the tunneling probability decreases exponentially with increasing barrier height, the tunneled current from the spin direction with the larger bandgap will be greatly attenuated, leading to a highly spin-polarized tunneling current.\cite{Moodera2007,Bainsla2018} SFMs such as EuSe have been shown to produce nearly 100\% spin polarization at low temperature. \cite{Moodera1993,Moodera1988} However, these well-known SFMs have magnetic transition temperatures below 70 K, making them unsuitable for room temperature applications. Thus, it becomes necessary to explore SFMs with high Curie temperatures in order to expand this technology for useful applications outside the laboratory.

Ferromagnetic SFMs produce sizable fringing magnetic fields that can interfere with neighboring components in high-density electronic architectures. In order to alleviate this problem, it is advantageous to find SFMs that maintain the exchange splitting with minimal total magnetic moment. Fully compensated ferrimagnetic Heusler compounds satisfy these requirements, as the internal exchange can be large enough to produce a usable exchange splitting, yet the total moment of the unit cell is small or near zero. Such materials are generally not exactly compensated, so a small moment will exist in real materials that is sufficient to maintain the asymmetric band splitting. Integrating this type of low-moment tunneling barrier with nonmagnetic electrodes would lead to a spin injection device that is essentially \textit{nonmagnetic}.\cite{Fujita2017} 

Recently,  Galanakis \textit{et. al.} have predicted a class of Heusler SFMs that would be effective at room temperature.\cite{Galanakis2013,Ozdogan2013a} Their calculations show that CrVTiAl has a magnetic transition near 2000 K, and an exchange splitting of 2$\Delta$E\textsubscript{ex} $=$ 0.28 eV on top of an E\textsubscript{g}$=$ 0.32 eV majority bandgap (Fig. \ref{fig:SF_device}). In addition, CrVTiAl is predicted to be a fully compensated ferrimagnet, allowing the final SF device to create a spin-polarized current without producing significant fringing fields. These properties make high-quality CrVTiAl a promising candidate for a room temperature SFM, and the compound has been synthesized in bulk form. \cite{Stephen2016, arxiv}

\begin{figure}[t]
\includegraphics[width=8cm]{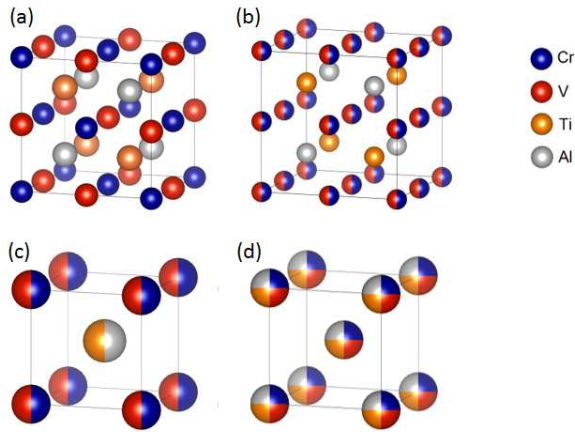}
\caption{Crystal structure of CrVTiAl with different degrees of atomic disorder. The disorder increases in going from the (a) fully ordered Y, (b) to the partially disordered L2\textsubscript{1}, (c) to the partially ordered B2, and finally (d) to the fully disordered A2 phase.}
\label{fig:struc}
\end{figure}

Given the difficulty of producing high-quality single-phase quaternary samples, especially as thin films suitable for device integration, it is essential to understand how disorder affects the electronic and magnetic structure.\cite{Jamer2013,Bainsla2016} Fig. \ref{fig:struc} illustrates some basic types of disorder in the quaternary Heusler structure. The fully-ordered Y-type structure is shown in Fig. \ref{fig:struc}(a). With L2\textsubscript{1} disorder, where the Cr and V atoms intermix, the symmetry is reduced to the familiar full-Heusler structure shown in Fig. \ref{fig:struc}(b). B2 disorder, shown in Fig. \ref{fig:struc}(c), leads to a smaller unit cell having a 2-atom basis with (Cr, V) atoms at the corners and (Ti, Al) in the center of the cubic cell. Finally, the A2 disorder, shown in Fig. \ref{fig:struc}(d) mixes all atoms equally and creates a simple body-centered cubic structure.  

In the present study we compare the experimental results on bulk polycrystalline CrVTiAl samples to parallel first-principles calculations of the magnetic and electronic structure,  where several types of disorder have been included. We show that the fully disordered A2 phase as well as the partially disordered B2 phase is metallic. Furthermore, Hall measurements indicate the onset of an added semiconducting contribution to the Hall coefficient and conductivity, originating from the presence of a small fraction of order (the fully-ordered Y or the partially disordered L2\textsubscript{1} state), which still preserves the bandgap. Results derived from both experiment and theory highlight the robustness of the electronic properties of CrVTiAl, confirming its suitability as an SFM. 

\section{Electronic structure of CrVTiAl}

\begin{figure*}
\centering
\includegraphics[width=0.95\textwidth]{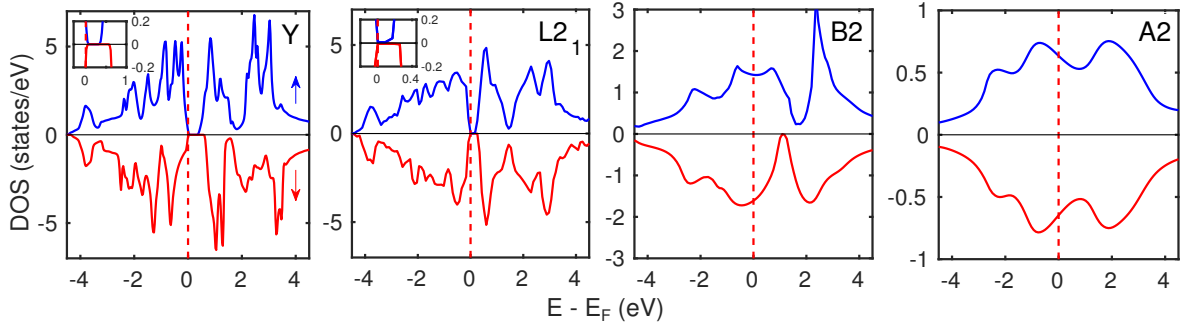}%{DOS_disorder_corrected}
\caption{DOS for various structures of CrVTiAl: fully ordered Y, partially ordered L2\textsubscript{1} and B2, and fully disordered A2 phases. The bandgap and spin polarization are maintained in the Y and L2\textsubscript{1} phases, but destroyed in the B2 and A2 phases.}
\label{fig:RT}
\end{figure*}

Disorder effects on the electronic structure were investigated using the all electron, fully charge- and spin-self-consistent Korringa-Kohn-Rostocker coherent-potential-approximation (KKR-CPA) first-principles scheme.\cite{PhysRevB.20.4025, PhysRevB.20.4035, PhysRevB.23.3608, PhysRevB.60.13396, bansil1993modern} The KKR-CPA assumes that atoms are immersed in an effective medium that is evaluated self-consistently to obtain the average Green’s function of the disordered system, and provides a realistic method for determining the electronic structure of randomly substituted alloys\footnote[1]{Supercell calculations for quaternary Heusler compounds have been reported by Neibecker et. al.\cite{PhysRevB.96.165131} to show that disorder effects are captured reasonably within the single-site KKR-CPA scheme.}. KKR-CPA does not involve any free parameters and thus is more satisfactory than other non-self-consistent approaches. \cite{arxiv}

The exchange-correlation contribution was incorporated within the local-spin-density approximation (LSDA) \cite{PhysRevB.23.5048} in the KKR-CPA computations. Charge- and spin-densities were converged to a high degree of accuracy. A muffin-tin radius of 2.51 a.u was used for all 4 atomic species. The experimental value of the lattice constant of \textit{a} = 6.136 \AA $\ $ was used (see discussion of XRD measurements below). In order to validate our simulations, we also modeled the Y-ordered phase using the Perdew-Burke-Ernzerhof (PBE) functional \cite{PhysRevLett.77.3865} as implemented in WIEN2k\cite{WIEN2k,Disclaimer} with a k-mesh of 18x18x18. Good agreement between the PBE and LSDA schemes is found, see Supplementary Material for details.

Effects of disorder were considered for three different types of random substitutions as follows. (1) Cr with V (L2\textsubscript{1}), (2) Simultaneous substitution of Cr with V and Ti with Al in the B2 structure and, (3) all four atomic species were equally mixed in the A2 phase (Fig. \ref{fig:struc}).  Fig. \ref{fig:RT} illustrates the densities of states (DOSs) of CrVTiAl in these four possible phases. Electronic properties are observed to evolve with disorder as follows. It can be seen from Fig. \ref{fig:RT} that the DOS of the fully ordered Y-structure is gapped near the Fermi level for both spins, with bandgaps of 0.34 eV and 0.58 eV in the majority and minority spins, respectively, in close agreement with the values in the literature literature. \cite{Galanakis2013}

It is important to note that the disordered L2\textsubscript{1} structure preserves the bandgap for both spins. Nevertheless, a general broadening of the features in the DOS is observed, thus reducing the gaps to values of 0.17 eV for the spin-up states and 0.25 eV for the spin-down states. This shrinkage could be caused by the presence of chemical disorder that disturbs the periodicity of the lattice; it could also be attributed to the fact that the LSDA exchange-correlation functional is known to underestimate the bandgap. \cite{QUA:QUA560280846} Moreover, we find that mixing of Y and L2\textsubscript{1} structures produces a robust SFM. As shown in Fig. 3, disorder effects close the gap in B2 and A2 structures. In particular, the electronic structure is sensitive to Ti-Al mixing, which is found to close the gap and produce a metal.

\begin{figure}
\includegraphics[width=.47\textwidth]{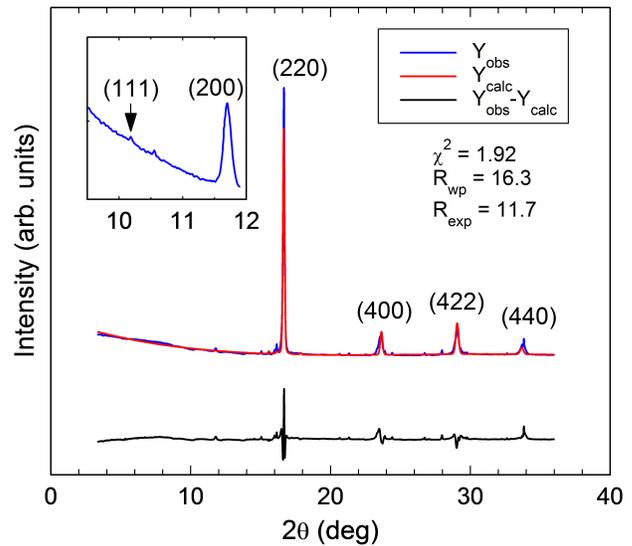}
\caption{Synchrotron XRD spectrum of CrVTiAl. The strong (220), (400), (422) and (440) reflection peaks have the same relative intensities for the fully ordered and fully disordered structures. (Inset) The intensity of the (200) peak is used to quantify the degree of B2 order. The weak (111) peak indicates the presence of a small percentage of an L2\textsubscript{1} ordered phase.}
\label{fig:XRD}
\end{figure}

The Y and L2\textsubscript{1} phases are clearly necessary for possible spin-filter applications, since they support bandgaps of spin-up and spin-down states of finite but unequal magnitudes. The fact that CrVTiAl remains an SFM even when mixing between Cr and V is introduced in the system, speaks to the robustness of the bandgaps, indirect for minority spins and direct for majority spins. The origin of the gaps lies in the bonding t\textsubscript{2g} and the non-bonding t\textsubscript{1u} hybridized states, as reported by Galanakis \cite{Galanakis2014}, with more detailed explanations provided in the work of {\"{O}}zdo{\u{g}}an. \cite{Ozdogan2013a}

DFT studies \cite{Galanakis2014,Ozdogan2013a,Galanakis2013} have shown that the \textit{p-d} hybridization is responsible for the fully compensated ferrimagnetism, with prominent magnetic moments on the Cr and V atoms. These magnetic properties are confirmed by our first-principles calculations. Our spin-polarized self-consistent calculation of the Y ordered phase using WIEN2k, yielded a total magnetic moment of 0 $\mu_B$/formula unit, consistent with the experimental results \cite{Stephen2016}, with the  predicted moments of: -2.63 $\mu_B$, 2.06 $\mu_B$, 0.40 $\mu_B$ and 0.02 $\mu_B$  on the Cr, V, Ti and Al sites, respectively, in good agreement with the literature. \cite{Galanakis2014}

\section{Experimental Results}
Bulk ingots were synthesized by arc melting stoichiometric quantities of elements followed by annealing for one week at 1000\degree C. Details of the synthesis can be found in Ref.  \cite{Stephen2016} Here, to further investigate the degree of atomic mixing in the Heusler lattice, synchrotron X-ray diffraction (XRD) was performed at the Cornell Energy Synchrotron Source A2-beamline ($\lambda$ = 0.6277 \AA). Fig. \ref{fig:XRD} shows the synchrotron XRD spectrum of the bulk CrVTiAl. The four strong XRD peaks corresponding to the (220), (400), (422), and (440) reflections are found to have the ideal intensity ratios common to the fully ordered Y-type structure, as well as the three disordered structures (L2$_1$, B2, and A2) illustrated in Fig. \ref{fig:struc}. There are also weak peaks corresponding to the (200) and (111) Bragg reflections, shown in the inset of Fig. \ref{fig:XRD}, which can be used to estimate the amount of chemical ordering in the phase.

\begin{figure*}[t]
\includegraphics[width=0.9\textwidth]{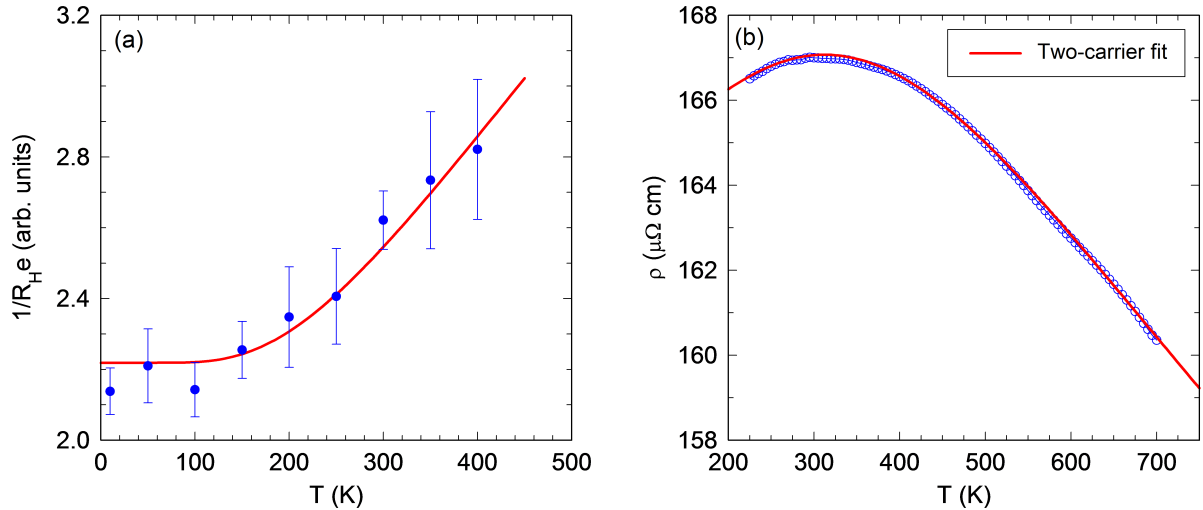}
\caption{(a) Reciprocal of the Hall coefficient versus temperature for the CrVTiAl bulk sample. The increase in 1/R\textsubscript{H} indicates an increasing carrier concentration with increasing temperature, however, additional linear-in-field contributions make the absolute concentration uncertain. \cite{Jamer2013,Stephen2016}(b) Resistivity of CrVTiAl with temperature. The resistivity reaches a maximum around 300 K before decreasing as a result of thermally activated carriers from the semiconducting L2\textsubscript{1} or Y-type phases.} 
\label{fig:XPort}
\end{figure*}

In order to determine the phase fraction of other structures contributing to the XRD pattern, Rietveld analysis employing FullProf software was used to refine the XRD data.\cite{Rodriguez-Carvajal1993,Roisnel2001} 
The bulk sample was found to be composed of 87 $\pm$ 5 wt\% of phases of the CrVTiAl compound with lattice constant \textit{a} = 6.136 $\pm$ 0.009 \AA. Pure elemental phases were found to be present, including Cr ($\approx$ 3 $\pm$ 1 \%) and Ti (10 $\pm$ 4 \%). It is possible, however, that other binary and ternary phases may also be present but are not detected due to low phase fraction. The XRD Bragg peak widths were used to estimate the coherent particle size. Williamson Hall analysis of the (220) and (440) peaks gave a particle size of $\tau = 20 \pm 4$ nm with a strain of $\epsilon = 2.5 \pm 0.3 \%$. \cite{Lewis2001} 

As demonstrated by Takamura \textit{et al.} \cite{Takamura2009}, the ratios of the observed (200) and (400) peak intensities ($I^{obs}$) to the expected peak intensities for the pure B2 phase ($I^{B2}$) gives the percentage of B2 phase in the sample according to the relationship
\begin{equation}
S_{B2}^2 =\frac{ I^{obs}_{200} / I^{obs}_{400}}{I^{B2}_{200}/I^{B2}_{400}},
\end{equation}
where $S_{B2}$ is the fraction of CrVTiAl that is in the B2 phase.
The integrated intensity ratios of the (200) and (400) peaks yields $S_{B2} = 0.58 \pm 0.02$, indicating that the structure is composed primarily of the B2 structure with a small admixture of the A2 structure.

A distinct (111) Bragg peak can be seen in the XRD data highlighted in the inset of Fig. \ref{fig:XRD}, but its weak intensity and characteristic background makes it difficult to obtain good quality Rietveld fits. Nevertheless, a comparison of the intensity of the (111) peak to that of the (220) peak gives an estimate of 6\% for the amount of L2\textsubscript{1} in the sample. We hypothesize that the presence of texture in the sample causes the L21 phase fraction to be underestimated. The predicted XRD patterns for the L2\textsubscript{1} and Y phases in this composition are differentiated only by small differences in the intensities of the (111) and (200) peaks due to the similarity in atomic form factors for the Cr and V atoms.

Notably, in addition to the substantial linear magnetic component arising from the compensated ferrimagnetism, there may be an additional diamagnetic contribution from the elemental phases (Cr, Ti, Al). However, the total measured mass susceptibility of $4.6 \times 10^{-5}  cm^3/g$ is too large to originate from the low concentrations of these elemental phases, indicating that the majority of this linear-in-field moment must arise from the CrVTiAl phase.\cite{footnote3}

The transport experiments presented here reveal strong evidence of thermally-activated semiconducting carriers, as verified by both Hall and resistivity measurements. We have previously reported metallic resistivity in this compound for low T that deviated from linearity in the range 200 K < T < 320 K, and attributed that behavior to a two-carrier model. \cite{Jamer2013, Stephen2016} The inverse Hall coefficient in Fig. \ref{fig:XPort} is seen to increase for increasing temperature as a result of this thermal activation, which clearly indicates an increase in the number of carriers at higher temperatures. However, this increase is only a qualitative measure as the Hall voltage could have additional linear-in-field contributions such as a temperature-independent anomalous Hall effect.

We have measured the zero-field resistivity the zero-field resistivity has been measured up to 700 K at the National High Magnetic Field Lab. The resistivity begins to decrease around 400 K and decreases rapidly for temperature up to 700 K. The high-temperature resistivity data is fit to a two-carrier model as detailed in Ref. \cite{Stephen2016}, where resistivity is modeled by considering two parallel and independent conduction channels, one with a constant number of carriers (metallic) and the other with thermally activated carriers (semiconducting). A simple phonon-dominated mobility is assumed for both channels. A fit to this model gives an activation energy of $\Delta E = 0.13  \pm  0.02$ eV, which is comparable to the previously reported value. \cite{Stephen2016} For an intrinsic semiconductor, the activation energy is given by $\Delta E = E_g/2$, so the present experimental value $E_g =$  0.26 eV for the bandgap lies between the predicted majority carrier bandgaps of 0.17 and 0.34 eV for the L2\textsubscript{1} and Y-type phases.  This is about 1/3 of the gap in germanium, and substantially larger than the thermal energy at room temperature. Our observation of similar temperature-dependent changes in both the resistivity and Hall coefficient is best explained by the aforementioned two-carrier model, which involves a combination of a metallic and semiconducting component components.\cite{Jamer2013,Stephen2016} Thus, out electrical transport measurements support the coexistence of metallic A2 and B2 phases alongside the presence of a semiconducting L2\textsubscript{1} or Y-type phase in the sample.

\section{Conclusions}
Our \textit{ab initio} calculations show that the bandgap and spin-polarization of CrVTiAl collapse under B2 and A2 disorder, but are preserved under L2\textsubscript{1} disorder. These predictions are consistent with the observed metallic-like electrical behavior that is dominant below T = 200 K, which we associate with B2- and A2-type disorder. However, the Hall and conductivity measurements indicate a semiconducting-like behavior above T = 200 K arising from thermally-activated carriers resulting from a small percentage of L2\textsubscript{1} ordered phase. The robustness of the computed spin-polarized bandgaps suggests that CrVTiAl can be an effective SFM, even without being fully chemically ordered in the Y phase, but it will benefit from improvements in its crystal structure.

\section{Supplementary Material}
The supplementary materials contain a comparison of the DOS for the ordered Y structure as calculated by KKR-CPA and WEIN2K. 

\section{Acknowledgements}
The XRD measurements were conducted at the Cornell High Energy Synchrotron Source (CHESS), which is supported by the National Science Foundation (NSF) under award DMR-1332208. We thank Jacob Ruff for his help with measurements taken at CHESS A2-beamline. Support for this project is acknowledged from the NSF, grant ECCS-0142738.The theoretical work was supported by the US Department of Energy (DOE), Office of Science, Basic Energy Sciences grant number DE-FG02-07ER46352, and benefited from Northeastern University's Advanced Scientific Computation Center (ASCC), and the NERSC supercomputing center through DOE grant number DE-AC02-05CH11231. A portion of this work was performed at the National High Magnetic Field Laboratory, which is supported by National Science Foundation Cooperative Agreement No. DMR-1644779 and the State of Florida. G. M. Stephen was supported in part by NSF IIP INTERN Program Supplement Award number 1601895.

\bibliographystyle{ieeetr}
\bibliography{References}

\begin{thebibliography}{10}

\bibitem{Graf2016a}
T.~Graf, C.~Felser, and S.~S.~P. Parkin, ``{Heusler Compounds: Applications in
  Spintronics},'' in {\em Handbook of Spintronics}, pp.~335--364, Springer,
  2016.

\bibitem{Wolf2001}
S.~A. Wolf, D.~D. Awschalom, R.~A. Buhrman, J.~M. Daughton, S.~von
  Moln{\'{a}}r, M.~L. Roukes, A.~Y. Chtchelkanova, and D.~M. Treger,
  ``{Spintronics: a spin-based electronics vision for the future.},'' {\em
  Science}, vol.~294, no.~5546, pp.~1488--95, 2001.

\bibitem{Moodera2010}
J.~S. Moodera, G.~X. Miao, and T.~S. Santos, ``{Frontiers in spin-polarized
  tunneling},'' {\em Phys. Today}, vol.~63, no.~4, pp.~46--51, 2010.

\bibitem{Moodera2007}
J.~S. Moodera, T.~S. Santos, and T.~Nagahama, ``{The phenomena of spin-filter
  tunneling},'' {\em J. Phys. Condens. Matter}, vol.~19, no.~16, p.~165202,
  2007.

\bibitem{Bainsla2018}
L.~Bainsla, K.~Z. Suzuki, M.~Tsujikawa, H.~Tsuchiura, M.~Shirai, and
  S.~Mizukami, ``{Magnetic tunnel junctions with an equiatomic quaternary
  CoFeMnSi Heusler alloy electrode},'' {\em App. Phys. Lett.}, vol.~112, no.~5,
  2018.

\bibitem{Moodera1993}
J.~S. Moodera, R.~Meservey, and X.~Hao, ``{Variation of the electron-spin
  polarization in EuSe tunnel junctions from zero to near 100{\%} in a magnetic
  field},'' {\em Phys. Rev. Lett.}, vol.~70, no.~6, pp.~853--856, 1993.

\bibitem{Moodera1988}
J.~S. Moodera, X.~Hao, G.~A. Gibson, and R.~Meservey, ``{Electron-Spin
  Polarization in Tunnel Junctions in Zero Applied Field with Ferromagnetic EuS
  Barriers},'' {\em Phys. Rev. Lett.}, vol.~61, no.~5, pp.~637--640, 1988.

\bibitem{Fujita2017}
H.~Fujita, ``{Field-free, spin-current control of magnetization in
  non-collinear chiral antiferromagnets},'' {\em Phys. Status Solidi - R},
  vol.~11, no.~4, pp.~1--5, 2017.

\bibitem{Galanakis2013}
I.~Galanakis, K.~{\"{O}}zdo{\u{g}}an, and E.~{\c{S}}a{\c{s}}{\i}o{\u{g}}lu,
  ``{A proposal for an alternative class of spin filter materials:
  Hybridization-induced high-TC ferromagnetic semiconductors CoVXAl (X=Ti, Zr,
  Hf)},'' {\em Appl. Phys. Lett.}, vol.~103, no.~14, p.~142404, 2013.

\bibitem{Ozdogan2013a}
K.~{\"{O}}zdo{\u{g}}an, E.~{\c{S}}a{\c{s}}{\i}o{\u{g}}lu, and I.~Galanakis,
  ``{Slater-Pauling behavior in LiMgPdSn-type multifunctional quaternary
  Heusler materials: Half-metallicity, spin-gapless and magnetic
  semiconductors},'' {\em J. Appl. Phys.}, vol.~113, no.~19, p.~193903, 2013.

\bibitem{Stephen2016}
G.~M. Stephen, I.~McDonald, B.~Lejeune, L.~H. Lewis, and D.~Heiman,
  ``{Synthesis of low-moment CrVTiAl: A potential room temperature spin
  filter},'' {\em Appl. Phys. Lett.}, vol.~109, no.~24, p.~242401, 2016.

\bibitem{arxiv}
Y.~Venkateswara, S.~Gupta, S.~S. Samatham, M.~R. Varma, Enamullah, K.~G.
  Suresh, and A.~Alam, ``{Competing magnetic and spin-gapless semiconducting
  behavior in fully compensated ferrimagnetic CrVTiAl: Theory and
  experiment},'' {\em Phys. Rev. B}, vol.~97, no.~5, p.~054407, 2018.

\bibitem{Jamer2013}
M.~E. Jamer, B.~A. Assaf, T.~Devakul, and D.~Heiman, ``{Magnetic and transport
  properties of Mn2CoAl oriented films},'' {\em Appl. Phys. Lett.}, vol.~103,
  no.~14, pp.~6--11, 2013.

\bibitem{Bainsla2016}
L.~Bainsla and K.~G. Suresh, ``{Equiatomic quaternary Heusler alloys: A
  material perspective for spintronic applications},'' {\em Applied Physics
  Reviews}, vol.~3, no.~3, p.~031101, 2016.

\bibitem{PhysRevB.20.4025}
A.~Bansil, ``{Coherent-potential and average $t$-matrix approximations for
  disordered muffin-tin alloys. I. Formalism},'' {\em Phys. Rev. B}, vol.~20,
  pp.~4025--4034, 1979.

\bibitem{PhysRevB.20.4035}
A.~Bansil, ``{Coherent-potential and average $t$-matrix approximations for
  disordered muffin-tin alloys. II. Application to realistic systems},'' {\em
  Phys. Rev. B}, vol.~20, pp.~4035--4043, 1979.

\bibitem{PhysRevB.23.3608}
A.~Bansil, R.~S. Rao, P.~E. Mijnarends, and L.~Schwartz, ``{Electron momentum
  densities in disordered muffin-tin alloys},'' {\em Phys. Rev. B}, vol.~23,
  p.~3608, 1981.

\bibitem{PhysRevB.60.13396}
A.~Bansil, S.~Kaprzyk, P.~E. Mijnarends, and J.~Tobola, ``{Electronic structure
  and magnetism of $Fe_{3x}V_{x}X$ (X=Si,Ga, and Al) alloys by the KKR-CPA
  method},'' {\em Phys. Rev. B}, vol.~60, pp.~13396--13412, 1999.

\bibitem{bansil1993modern}
A.~Bansil, ``Modern band theory of disordered alloys: basic concepts including
  a discussion of momentum densities,'' {\em Zeitschrift f{\"u}r Naturforschung
  A}, vol.~48, no.~1-2, pp.~165--179, 1993.

\bibitem{PhysRevB.96.165131}
P.~Neibecker, M.~E. Gruner, X.~Xu, R.~Kainuma, W.~Petry, R.~Pentcheva, and
  M.~Leitner, ``Ordering tendencies and electronic properties in quaternary
  heusler derivatives,'' {\em Phys. Rev. B}, vol.~96, p.~165131, Oct 2017.

\bibitem{PhysRevB.23.5048}
J.~P. Perdew and A.~Zunger, ``Self-interaction correction to density-functional
  approximations for many-electron systems,'' {\em Phys. Rev. B}, vol.~23,
  pp.~5048--5079, 1981.

\bibitem{PhysRevLett.77.3865}
J.~P. Perdew, K.~Burke, and M.~Ernzerhof, ``Generalized gradient approximation
  made simple,'' {\em Phys. Rev. Lett.}, vol.~77, pp.~3865--3868, 1996.

\bibitem{WIEN2k}
P.~Blaha, K.~Schwarz, G.~Madsen, D.~Kvasnicka, and J.~Luitz, ``{WIEN2k, An
  Augmented Plane Wave + Local Orbitals Program for Calculating Crystal
  Properties},'' 2001.

\bibitem{Disclaimer}
{{Any mention of commercial products is for information only; it does not imply
  recommendation or endorsement by NIST}}.

\bibitem{QUA:QUA560280846}
J.~P. Perdew, ``Density functional theory and the band gap problem,'' {\em Int.
  J. Quantum Chem.}, vol.~28, no.~S19, pp.~497--523, 1985.

\bibitem{Galanakis2014}
I.~Galanakis, {\"{O}}zdo{\u{g}}an, and E.~{\c{S}}a{\c{s}}{\i}o{\u{g}}lu,
  ``{High-TC fully compensated ferrimagnetic semiconductors as spin-filter
  materials: the case of CrVXAl (X = Ti, Zr, Hf) Heusler compounds},'' {\em J.
  Phys. Condens. Matter}, vol.~26, no.~8, p.~086003, 2014.

\bibitem{Rodriguez-Carvajal1993}
J.~Rodr{\'{i}}guez-Carvajal, ``{Recent advances in magnetic structure
  determination by neutron powder diffraction},'' {\em Phys. B Condens.
  Matter}, vol.~192, pp.~55--69, 1993.

\bibitem{Roisnel2001}
T.~Roisnel and J.~Rodr{\'{i}}quez-Carvajal, ``{WinPLOTR: A Windows Tool for
  Powder Diffraction Pattern Analysis},'' {\em Mater. Sci. Forum},
  vol.~378-381, no.~December, pp.~118--123, 2001.

\bibitem{Lewis2001}
L.~H. Lewis, A.~R. Moodenbaugh, D.~O. Welch, and V.~Panchanathan, ``{Stress,
  strain and technical magnetic properties in `exchange-spring' Nd2Fe14B +
  $\alpha-Fe$ nanocomposite magnets},'' {\em J. Phys. D. Appl. Phys.}, vol.~34,
  no.~5, pp.~744--751, 2001.

\bibitem{Takamura2009}
Y.~Takamura, R.~Nakane, and S.~Sugahara, ``{Analysis of L21-ordering in
  full-Heusler Co2FeSi alloy thin films formed by rapid thermal annealing},''
  {\em J. Appl. Phys.}, vol.~105, no.~7, p.~07B109, 2009.

\bibitem{footnote3}
{The saturating moment observed by Venkateswara et. al with T\textsubscript{c}
  = 800 K was initially observed in this sample as well, however the saturating
  moment disappeared after sanding removing a yellow oxide from the surface of
  the sample. It was thus attributed to this oxide rather than the bulk
  CrVTiAl. The positive linear component continued to increase up to 1000 K,
  indicating that the compensated ferrimagnetic component has
  T\textsubscript{c} > 1000 K.}

\end{thebibliography}
%\begin{thebibliography}{10}

%\end{thebibliography}

\end{document}